\documentstyle[12pt,aasms4]{article}

\lefthead{Afanasiev and Sil'chenko}
\righthead{Global Structure and Kinematics of NGC 2841}

\begin{document}

\title{Global Structure and Kinematics of the Spiral Galaxy NGC~2841}

\author{V. L. Afanasiev}
\affil{Special Astrophysical Observatory, Nizhnij Arkhyz,
    357147 Russia\\
    Electronic mail: vafan@sao.ru}

\and

\author{O. K. Sil'chenko\altaffilmark{1}}
\affil{Sternberg Astronomical Institute, Moscow, 119899 Russia\\
       Isaac Newton Institute, Chile, Moscow Branch\\
     Electronic mail: olga@sai.msu.su}

\altaffiltext{1}{Guest Investigator of the RGO Astronomy Data Centre}

\begin{abstract}

Investigation of gaseous and stellar kinematics and of broad-band
$VRI$ and narrow-band $H_\alpha$ and [NII]$\lambda6583$ images is
performed for the central part ($R<4$ kpc) of the regular spiral galaxy
NGC~2841. We have found emission-line splitting at $R<20\arcsec$ and
three-component LOSVD for the stars in the radius range
$6\arcsec \div 100\arcsec$. Morphological analysis reveals strong
narrow shock fronts close to the major axis in the radius range
of $30\arcsec \div 50\arcsec$, a turn of the isophote major axis
by 5\arcdeg\ and strongly negative Fourier coefficient $a_4$
(boxy isophotes) in the radius range of $15\arcsec \div 33\arcsec$.
In principle, all these features may be explained in the frame
of a triaxial bulge hypothesis.

\end{abstract}


\section{Introduction}

NGC~2841 is a regular early-type spiral galaxy, rather isolated,
without any morphological signs of past interaction. It has a weak
LINER nucleus (e.g., \cite{ho95}) and a flocculent spiral structure
(\cite{elm}). We have investigated the innermost part of the galaxy
inside $R=10\arcsec$ with the Multi-Pupil Field Spectrograph of
the 6m telescope (\cite{we97}). The stellar nucleus of NGC~2841
was found to be chemically decoupled, being more metal rich
than surrounding bulge by a factor of 2.3; moreover, nuclear
ionized gas demonstrated rotation in the plane strictly orthogonal
to the global galactic plane (and to the rotation plane of stars
in the center of the galaxy too). It looks like a small gaseous
polar ring. Global polar rings are usually treated as signatures
of past interaction with gas accretion or even of a minor merger.
But such an event cannot affect only the very center of the galaxy;
there must be any peculiarities in the global structure. We have found
some hints on the presence of a dynamically decoupled stellar
component in the bulge of NGC~2841 (\cite{we97}); but a further
investigation of the global structure and kinematics of the
galaxy was needed to understand what we see here.

NGC~2841 is known to possess a quite regular structure. The only
claim about its peculiarity was result of Varela et al.
(\cite{var96}) that the bulge of NGC~2841 is a triaxial ellipsoid
with varying axis ratio. Starting from the pioneer work of
Zaritsky \&\ Lo (\cite{zl86}), a triaxial bulge in a spiral galaxy
is usually detected by a turn of isophote major axis at the galactic
center. But the isophote major axis turn in NGC~2841, which causes
Varela et al. (\cite{var96}) to claim a presence of the triaxial
bulge, is quite small, only 10\arcdeg\ according to their
measurements ($P.A.=154\arcdeg$ for the bulge and $P.A.=144\arcdeg$
for the disk). If we take into account a lot of other independent
determinations of the line-of-nodes orientation in NGC~2841
(photometry: $P.A._0$=147\arcdeg, RC3, \cite{mor98},
$P.A._0$=150\arcdeg, \cite{hs96,we97}; neutral hydrogen kinematics:
$P.A._0$=150\arcdeg, \cite{rots}), this value may be even smaller,
4\arcdeg\--7\arcdeg, and then is comparable with the measurement
errors. Besides, the orientation of the isophote major axis
$P.A.$=154\arcdeg, which is ascribed to the triaxial bulge by
Varela et al. (\cite{var96}), is observed only in the radius range
of 20\arcsec--40\arcsec; closer to the center the $P.A.$ value falls
to the outermost level, $P.A.$=147\arcdeg, and Varela et al. are
forced to suggest a presence of ANOTHER inner bar aligned with the
line of nodes to explain a behaviour of isophote position angle
dependence on the radius. The structure begins to look too complex
for such a small position angle variation. So, the question of
triaxial bulge presence in NGC~2841 must be regarded more carefully
by involving kinematical data for gas and stars in the central
region dominating by the bulge potential. Our paper treats just
this problem. A description of observational data used for the
analysis is presented in Sec.~2. The kinematical results are
discussed in Sec.~3, and the morphological results -- in Sec.~4.
Finally, in Sec.~5 we briefly discuss some details and conclude
with a summary of our study.

\section{Observational Data}

The data which we analyse in this work include long-slit spectra
taken almost along the major axis of NGC~2841, broad-band $VRI$
panoramic photometry and narrow-band images obtained through
the scanning Fabry-Perot interferometer in combination with
interference filters centered on the redshifted emission lines
$H_\alpha$ and [NII]$\lambda$6583. We have used facilities
of the La Palma Archive
as well as our own observations made at the 6m telescope of the
Special Astrophysical Observatory (Nizhnij Arkhyz, Russia).

Spectral data were obtained on January 6 and 7, 1991, at the
4.2m William Herschel Telescope on La Palma with the ISIS, red
arm, equipped by a CCD $800 \times 1180$. Two spectral ranges were
exposed, 5830--6680 \AA\ included the strongest emission lines
$H_\alpha$ and [NII]$\lambda$6583 and 8000--8840 \AA\ included
the strongest stellar absorption lines \ion{Ca}{2}$\lambda$8498,
8542, 8662. The dispersion was 0.74~\AA\ per pixel which corresponds
to the spectral resolution of 2~\AA, the slit width was $0\farcs9$,
and a scale along the slit was $0\farcs 335$ per pixel. The galaxy
was exposed in two position angles, $P.A.=145\arcdeg$ and 153\arcdeg,
both close to the major axis. Exposure times were set equal to
1 hour, except the exposure in $P.A.=153\arcdeg$, 5830--6680 \AA,
which was of 40 min. Copper-neon and copper-argon comparison lamps
were used for wavelength calibration; a check by measuring night-sky
emission lines has shown that the inner accuracy of a single velocity
measurement inside one spectrum frame is not worse than 2 km/s,
though a systematic shift of the whole frame by 10--30 km/s is
possible between neighboring exposures. Two bright stars,
HR~3125 (K1III) and HR~3222 (G8III), were used as templates for
a cross-correlation in the spectral range of 8000-8840 \AA\
to obtain stellar velocities along the major axis of NGC~2841.

The photometric data were obtained on May 6, 1988, at the 1m Jacobus
Kapteyn Telescope on La Palma with CCD $385 \times 578$, through
the broad-band filters $VRI$. A spatial scale was $0\farcs 30$ per
pixel, so only the central part of the galaxy was exposed; the
seeing quality ($FWHM$) was $0\farcs 7$--$0\farcs 8$. The exposure
times were 300+300 sec ($V$), 300+200 sec ($R$), and 200+200 sec
($I$). No photometric standards were observed, so we have not made
any attempts to calibrate these CCD frames. Instead we have made
an isophotal analysis deriving radial dependencies of the major-axis
position angle, isophote ellipticity, and boxiness.

At the 6m telescope we have observed NGC~2841 with the scanning
Fabry-Perot interferometer (\cite{dodo}) twice: in December 1996
in the emission line $H_\alpha$ and in May 1997 in the emission line
[NII]$\lambda$6583. The observations were undertaken in the frame
of the project of Prof. A. M. Fridman dedicated to a search of
large-scale vortices in gas velocity fields of spiral galaxies.
Here we present only narrow-band images of NGC~2841 in $H_\alpha$
and [NII] emission lines obtained as by-products of these
observations. A CCD detector with format of $1060 \times 1180$
was used with a binning of $2 \times 2$; a resulting scale of
$0\farcs 455$ per pixel after binning $2 \times 2$ allowed to
register an area of $4\arcmin \times 4\farcm 5$. The seeing quality
was $2\farcs 3$ in December 1996 and $1\farcs 7$ in May 1997.

All the observational data were reduced by using the software
developed in the Special Astrophysical
Observatory (\cite{vlas}) and the software ADHOC developed in
the Marseille Observatory (\cite{boul}).

\section{Kinematical Results}

\subsection{Gas line-of-sight velocity distributions}

The long-slit spectra obtained in the spectral range of 5830--6680
\AA\ reveal a presence of quite measurable emission lines
$H_\alpha$ and [NII] in the full radius range of 0\arcmin--2\arcmin\
covered by the observations. Inside $R=60\arcsec$ the nitrogen
emission line $\lambda$6583 is almost everywhere stronger than
$H_\alpha$ implying a shock excitation mechanism; sometimes
the emission lines are clearly splitting with a velocity separation
of the components up to 150 km/s. We have measured gas line-of-sight
velocities along the slit by searching for line baricenters and by
applying a Gauss analysis to the multi-component profiles; the
complete results are presented in Fig.~1, and the results for the
central part are presented in Fig.~2. Outside $R\approx 50\arcsec$
the measured velocity distributions are quite regular, the rotation
curve looks flat, in accordance with numerous previous investigations,
at the level of $v_{rot}=313$ km/s assuming after Whitmore et al.
(\cite{wrf84}) galaxy inclination to be 65\arcdeg.
In the inner part of NGC~2841, particularly inside
$R\approx 20\arcsec$, the behaviour of the gas line-of-sight
velocities is so complex that we cannot explain all the details of
the velocity profiles. But
some typical features of the distributions can be interpreted as
evidences for a bar-like potential. First of all, though the
cross-sections in $P.A.=145\arcdeg$ and in $P.A.=153\arcdeg$ are
positioned very tightly, Fig.~2a and Fig.~2b look quite different.
It means that the structure in the center of NGC~2841 which causes
the emission line splitting is very subtle, almost one-dimensional.
Though the global disk of NGC~2841 is moderately inclined to the line
of sight, by some
60\arcdeg--65\arcdeg\ (Varela et al. \cite{var96}, \cite{mor98}),
it seems that we see a bar aligned with the line of nodes through
the rotating gaseous disk. In $P.A.=153\arcdeg$ (Fig.~2a) we can
note two straight segments of the velocity distributions separated
by 150--180 km/s in the same radius range, namely,
10\arcsec--20\arcsec\ to the north from the center. It looks like
a simultaneous registration of a fast-rotating gaseous disk and
of a slow-rotating bar edge which is traced by gas radiating
in a shock wave. A similar emission-line splitting has been
predicted in the dynamical work of Kuijken and Merrifield (\cite{km95})
who have considered visible effect of edge-on bars on major-axis
line-of-sight velocity profiles.
In $P.A.=145\arcdeg$ (Fig.~2b) two straight
segments have an opposite slope and are located in the radius
ranges of 0\arcsec--10\arcsec\ to the south and 10\arcsec--20\arcsec\
to the north from the center. This configuration with a visible
velocity break in the center of the symmetry can be understood
if we remind that there exists a radius range (usually between
two ILRs) where any bar provokes a predominance of gas orbits
elongated orthogonally to this bar (e. g. \cite{mulder}). The orbits
highly elongated along the line of sight (seen "end-on") would
provide just the configuration similar to that in Fig.~2b.
Solely, a position of the velocity distribution symmetry center
in this configuration
must be shifted by 5\arcsec\ to the north from the photometric
center. But it is quite possible: in our previous investigation
(\cite{we97}) made with the panoramic spectrograph providing
a two-dimensional velocity field we have already obtained
a position of the symmetry center of the gas velocity field
shifted to the north with respect to the brightness center position.

\subsection{Stellar line-of-sight velocity distributions}

We have derived stellar line-of-sight velocities by cross-correlating
sky-  and continuum-subtracted near-infrared galactic spectra
line-by-line with the spectra of the template stars observed the same
night. An example of the cross-correlation peaks is presented
in Fig.~3. In the full radius range where the cross-correlation peaks
are measurable -- from 45\arcsec\ to the south from the nucleus
toward 100\arcsec\ to the north -- they demonstrate a
multi-component structure. Velocity differences between
the components are sufficiently large so extraction of the
components is a procedure easy enough. We have used a Gauss analysis
for this purpose. Curiously, as it is illustrated in Fig.~3a, to
the south from the nucleus the intermediate-velocity component
is the strongest one, and to the north from the nucleus, at
$R > 25\arcsec$, the extremely blueshifted component becomes the
most prominent. This asymmetry is the first sign of a triaxial
bulge. Unfortunately, we have not been able to trace
stellar velocity distributions toward the center of NGC~2841:
inside $R\approx 6\arcsec$ stellar velocity dispersion strongly
increases (or additional kinematical components arise?), and an
unambiguous component separation becomes impossible. Beyond
$R\approx 6\arcsec$ all the cross-correlation peaks can be decomposed
mainly into three components; the corresponding three kinematical
subsystems are presented in Fig.~4. Attached error bars
characterize differences between velocity determinations by using
two different template stars.

Unlike the gaseous velocity distributions in Fig.~1, the stellar
velocity distributions in two different position angles, 145\arcdeg\
and 153\arcdeg, are quite identical. In Fig.~4 we see three stellar
kinematical subsystems -- two prograde and one retrograde. Among
two prograde subsystems, one rotates by a factor of 3 faster than
another. The rotation velocities of the slow-rotating prograde system
and of the retrograde system are comparable. We have called these
three subsystems
"direct bulge", "disk", and "retrograde bulge". The rotation velocity
of the disk is practically equal to the rotation velocity of the
gaseous disk, so we conclude that it is a dynamically cold stellar
subsystem. The semi-amplitudes of velocity variations for the both
"bulges" are of order of 100 km/s; and we cannot prove that the
rotation of these components is circular. In the context of a
triaxial bulge hypothesis an existence of two "bulge" kinematical
subsystems is easily explained if the bulge is triaxial,
seen edge-on, and slightly tumbling; in this case we see streaming
stellar motions along the bar simultaneously in both directions.

\section{Morphological Signatures of the Bar}

\subsection{Narrow-band $H_\alpha$ and [NII] images}

Figure~5 presents isophotes of the surface brightness distributions
in the emission lines $H_\alpha$ (a) and [NII]$\lambda$6583 (b)
for the very center of NGC~2841 $30\arcsec \times 30\arcsec$.
Earlier we have already constructed an analogous map in [NII] by
using the observational data obtained with the Multi-Pupil Field
Spectrograph of the 6m telescope (\cite{we97}); an agreement
of the Fig.~5b presented here with the Fig.~11 in the previous
work (\cite{we97}) is perfect even in details: again we see
that the overall intensity distribution is aligned with the
line of nodes ($P.A._0=150\arcdeg$), again we can notice an
extra-component to the north from the nucleus aligned in
$P.A.\approx 50\arcdeg$, and even splitting southern "tail"
of the surface brightness distribution is exactly reproduced.
Interestingly, the $H_\alpha$ image of the center of NGC~2841
(Fig.~5a) has appeared to be quite different: it is much more
symmetric than the [NII]$\lambda$6583 image though it
demonstrates also some small peculiarities, such as concaves
of the outer isophote at the $P.A.\approx 70\arcdeg$ (dust lane?).
A lack of resemblance between the two images promises something
interesting if mapping [NII]-to-$H_\alpha$ emission line ratio.

Figure~6 presents a full-format map of [NII]$\lambda$6583 to
$H_\alpha$ flux ratio with superimposed isophotes of the $H_\alpha$
image. The map is normalized to the value [NII]/$H_\alpha$=2
in the nucleus which was obtained from the La Palma long-slit
spectra. First of all, we can notice that in the spiral arms
which are clearly sketched by the $H_\alpha$ emission maxima
the [NII] to $H_\alpha$ ratio falls below 0.5. It is consistent
with an excitation of the ionized gas by hot massive stars.
But besides the spiral arms, which are distinct by the low
[NII]-to-$H_\alpha$ ratio, there are also two long, slightly
curved loci, which are distinct by a high [NII]-to-$H_\alpha$
ratio. These features are located in the radius range of
30\arcsec--50\arcsec; their [NII]/$H_\alpha$ values between 3 and 4
allow to regard them as shock wave fronts. Their morphology
resembles that of dust lanes which are often seen on the edges
of bars and which are also related to shock wave fronts. Another
evidence for a shock origin of the high [NII]/$H_\alpha$ ratio
regions is presented in Fig.~7. The slit at the position angle
$P.A.=153\arcdeg$ crossed the southern feature at
$R\approx 40\arcsec$ and $R\approx 50\arcsec$. Figure~7 shows
a corresponding piece of the gas velocity curve. The nitrogen
emission line in the radius range of $-54\arcsec \div -38\arcsec$
has a double-peaked profile. We have fitted it by two Gaussians;
one of them has given velocities corresponding to the regularly
rotating disk, the second -- the velocities smaller by some 100 km/s.
This velocity splitting is a sure signature of shock waves.

\subsection{Broad-band $VRI$ photometry}

A detailed surface photometry through the broad-band filters was
already performed for NGC~2841 by Varela et al. (\cite{var96}).
They have used CCD observations from the 4.2m WH Telescope of La
Palma taken under good seeing conditions. The data presented here
are obtained at the much smaller telescope, but also under very
good seeing conditions, so the quality of our morphological
results has appeared
to be not worse than in their work. Figure~8 presents radial
dependencies of isophote major axis position angle, of isophote
ellipticity, and of the fourth Fourier coefficient $a_4$ called
"boxiness". The former two dependencies at $a > 5\arcsec$ agree
perfectly with the results of Varela et al. (\cite{var96});
the boxiness was not analysed by them. The major axis position
angle shows small variations over the full radius range under
consideration; the central increase of $P.A.$ toward 156\arcdeg\
must be real because of agreement with the results of our analysis
of HST/WFPC image for NGC~2841 (\cite{we97}). Besides the $P.A.$
increase in the very center, we have also a plateau of $P.A.$
in the radius range of 13\arcsec--35\arcsec\ at the level of
$P.A.\approx 154\arcdeg$; outside this radius range the isophotes
are aligned with the line of nodes. Varela et al. (\cite{var96})
have obtained the same plateau in the same radius range and have
interpreted it as a signature of bulge triaxiality. We agree
because the same radius range is distinguished in Fig.~8c by
strongly negative $a_4$ values in the all three filters. It is
quite consistent with a picture of the large-scale bar seen
edge-on implied by the kinematics (see the previous Section),
because numerical studies of orbit instabilities have shown that
barred galaxies must have boxy bulges (e. g. \cite{p85}).
Inside $R\approx 12\arcsec$ the situation is not so clear:
the major axis position angle coincides with that of the line of
nodes, and $a_4$ is only mildly negative. Varela et al. (\cite{var96})
suggested the second, inner bar, but for this hypothesis there are
no arguments: their reference to Keel (\cite{keel}) is erroneous,
because Keel (\cite{keel}) claimed a one-sided GASEOUS bar in the
center of NGC~2841, not a two-sided stellar one. Perhaps it would
be more correct to classify a zone of $R$=4\arcsec--12\arcsec\ as
a transition from a slightly warped nuclear disk with a radius
of 2\arcsec\ to the (possibly torus-like?) triaxial bulge. A
behaviour of the isophote ellipticity (Fig.~8b) demonstrates
characteristic oscillations which may result from the presence
of two radially limited oblate components.

\section{Discussion and Conclusions}

The information which we have obtained on the kinematics of gas and
stars in NGC~2841 is rich and somewhat unexpected. The ionized-gas
emission lines are doubling mainly inside $R=20\arcsec$; but the stellar
LOSVD is multi-component up to $R=100\arcsec$! The stellar rotation
inside $R=40\arcsec$ in NGC~2841 was once investigated by Whitmore et al.
(\cite{wrf84}): two long-slit cross-sections, along $P.A.=150\arcdeg$
and along $P.A.=60\arcdeg$, were obtained under a spectral resolution
of 3.4 \AA, that is twice worse than ours, and no subsystem
multiplicity were noticed. As for gas rotation in this region
Whitmore et al. (\cite{wrf84}) said, with the reference to Rubin and
Thonnard (private communication), that the innermost measurable
emission was detected at $R=52\arcsec$, and for the inner disk
the rotation curve was extrapolated from this point to the center.
The mean major-axis stellar velocity distribution of Whitmore et al.
(\cite{wrf84}) showed a prominent minimum at $R=11\arcsec$; the
velocity decrease was measured to be steeper that a Keplerian one,
and Whitmore et al. (\cite{wrf84}) accepted a triaxiality of the
bulge. The bulge rotation velocity projected on the line of sight
was estimated by them (after subtracting the extrapolated disk)
as $70 \pm 30$ km/s. This estimate seems to be quite reasonable:
by the direct subsystem decomposition we have measured
$v_{rot}^{los}$=106 km/s for the "direct" bulge and 76 km/s for
the "retrograde" bulge.

Here we must mention that our direct measurement of the bulge rotation
curve is not unique. For example, Wagner et al. (\cite{wag89}) had
presented two parallel rotation curves for the disk and bulge of
NGC~4594 up to $R=40\arcsec$; they have noticed that correlation
peaks were doubling and, trying to decompose them with a lot of
precautions, they have obtained two stellar subsystems, both cold
enough, one of which rotated with the velocity of $\sim$ 300 km/s,
and the other -- slower by 120 km/s. The words "disk" and "bulge"
were pronounced in their Discussion. But in NGC~4594 there were
no any "counterrotating" subsystem which is seen, as third by its
luminosity, in NGC~2841.

A counterrotating subsystem was reported to be found in another
early-type spiral galaxy, NGC~7217, by Merrifield \& Kuijken
(\cite{mk94}). They have measured a double-peaked LOSVD and have
claimed a discovery of two counterrotating stellar disks in NGC~7217;
they argued that outside $R=10\arcsec$ a bulge contribution in this
galaxy is negligible and cannot affect the measured stellar kinematics.
But their photometric arguments were wrong: Buta et al.
(\cite{b95}) have decomposed a high-quality surface brightness profile
of NGC~7217 and have found that a de Vaucouleurs' bulge dominates
in this galaxy over the whole radius range. So perhaps in NGC~7217
Merrifield \& Kuijken (\cite{mk94}) saw just the same "prograde" and
"retrograde" bulges as we have seen in NGC~2841. Searching for a cause
of multi-ring structure in NGC~7217, Buta et al. (\cite{b95}) have
also detected a slight triaxiality of the galaxy bulge; it implies
that the hypothesis of elliptical stellar streams may be a quite
reasonable explanation of what Merrifield \& Kuijken (\cite{mk94})
saw in NGC~7217.

If we summarize the results of this work and those of our first
paper (\cite{we97}) on NGC~2841, we obtain a curious list of
kinematical and morphological features which must be jointed
in the frame of one hypothesis. All of them are schematically
shown in Fig.~9. These are:

\begin{itemize}

\item{the most prominent photometric appearance of a bar in the
radius range of 15\arcsec--33\arcsec;}
\item{a presence of strong shocks on the trailing sides of the
photometric bar in the radius range of 30\arcsec--50\arcsec;}
\item{in the radius range of 5\arcsec--15\arcsec\ gas is on orbits
in the galactic plane but elongated orthogonally to the bar;}
\item{inside $R\approx 5\arcsec$ gas rotates in the plane orthogonal
to the galactic plane (\cite{we97});}
\item{the visible center of gas rotation is shifted by
3\arcsec--5\arcsec\ to the north along the major axis
with respect to the brightness center (\cite{we97}, this paper);}
\item{star motions in the radius range of 4\arcsec--10\arcsec\
have a significant line-of-sight velocity component along the galaxy
minor axis -- either a polar ring or elongated plane orbits
orthogonal to the bar (\cite{we97});}
\item{in the radius range of 10\arcsec--100\arcsec\ we see two
"counterrotating" stellar bulges, moreover, to the north-west from
the center we see them both through the stellar disk -- the picture is
consistent with elliptical streaming of stars along the photometric
(but much shorter!) bar.}

\end{itemize}

During the last two decades there were a lot of papers on gas response
to an ovally distorted potential. Though the results seem to be model
dependent, however some things are firmly established. We know, for
example, that shocks at the trailing edges of a bar are a signature
of the low contrast of the oval potential distortion (\cite{m87}) and
that plane orthogonal orbit family dominates inside ILR
(e. g. \cite{mulder}). The most prominent theoretic bar manifestation
in the gaseous disk must be perhaps two-armed spiral pattern --
global density waves. Though NGC~2841 is a flocculent spiral galaxy,
recently two smooth (grand-design?) spiral arms have been detected
by Block et al. (\cite{bew96}) on the $K'$ (2.1$\mu$m) image of this
galaxy. The arms look dark; Block et al. (\cite{bew96}) suppose that
they are density waves in the dust-gaseous disk of the galaxy.
Therefore, these arms are just what must accompany the global bar
which we report in this paper. Interestingly, NGC~2841 is not the
only flocculent galaxy which demonstrates two-armed grand design
spirals in the $K'$ band; Thornley (\cite{thorn96}) has found them
in three flocculent galaxies of four ones which were investigated,
and Grosbol and Patsis (\cite{gp98}) report several more objects
of this kind.
One of the galaxies, NGC~5055, has been studied in detail by
Thornley and Mundy (\cite{thorn97}); the infrared spiral arms have
appeared to demonstrate also HI and CO concentrations and prominent
gas streaming motions. So, global density waves are perhaps
frequent enough in the gaseous disks of flocculent galaxies, and
nothing prevents flocculent galaxies to possess triaxial bulges.

Another feature of gas behaviour found by us -- nuclear
gaseous "polar ring" (\cite{we97}) -- is not predicted by
detailed dynamical simulations of bars
because these simulations are mostly two-dimensional, and here we
need to involve $z$ direction. But there was an interesting
observational note of Sofue and Wakamatsu (\cite{sw94}); they have
seen an asymmetrical dust lane orthogonal to the global bar in the
center of face-on SB galaxy M~83. Their physical arguments in favor
of polar orbit predominance for the nuclear gas of barred galaxies
are convincing enough.

Theoretical predictions of bar effects on stellar kinematics do not
contradict our results too. The two bulge kinematical subsystems,
prograde and retrograde, can be explained either by specific projection
of elliptical streamlines of stars in the triaxial potential or by a
presence of retrograde orbits in the triaxial potential: e. g., the
recent dynamical investigation of Wozniak and Pfenniger (\cite{wp97})
has shown that in different models from 15\%\ to 30\%\ (by mass) stars
of a bar are on retrograde orbits. Perhaps, a set of phenomena which
we have derived here for NGC~2841 is sufficient for specialists in
dynamics to build a model of the triaxial bulge in this galaxy and to
answer if we can explain the whole phenomena in the frame of this model
or some merger events must be involved.

\acknowledgements
We are grateful to Prof. A. M. Fridman for a permission to use the
observational data obtained in the frame of his observational
project prior to publication. We thank the astronomers of the Special
Astrophysical Observatory Drs. S. N. Dodonov, V. V. Vlasyuk, and A. N.
Burenkov for providing the observations at the 6m telescope with the
scanning Fabry-Perot interferometer, and we thank Drs. V. V. Vlasyuk
and J. Boulesteix from the Marseille Observatory for the possibility
to use their data reduction programs. We are also grateful to Dr. R.
Peletier who has attracted our attention to the observational data
on NGC~2841 which are kept in the La Palma Archive.
This research has made use of the La Palma Archive. The telescopes
WHT and JKT are operated on the island of La Palma by the Royal
Greenwich Observatory in the Spanish Observatorio del Roque de los
Muchachos of the Instituto de Astrofisica de Canarias.
The 6m telescope is operated under the financial support of
Science Department of Russia (registration number 01-43). The work
was supported by the grant of the Russian Foundation for Basic
Researches 98-02-16196 and by the Russian State Scientific-Technical
Program "Astronomy. Basic Space Researches" (the section "Astronomy").

\newpage

\figcaption{Radial distributions of ionized-gas line-of-sight
velocities in $P.A.=153\arcdeg$ ({\it a}) and in $P.A.=145\arcdeg$
({\it b}). A typical r. m. s. error of the velocities determined
by searching for emission-line baricenters is 2 km/s, errors of
the Gauss-analysis determinations are shown by vertical bars}

\figcaption{The same as Fig.~1, but only for the central part
of NGC~2841}

\figcaption{{\it a} -- Examples of correlation peaks from
cross-correlation of galactic spectra in the spectral range of
8000--8840 \AA \AA\ with the spectra of template stars; two
profiles, at $r=-40\arcsec$ (positive velocity shift) and
at $r=+40\arcsec$ (negative velocity shift), are presented
for the case $P.A.=145\arcdeg$ and template star HR~3125;
{\it b} -- two autocorrelation peaks, HR~3125 {\it vs} HR~3125
and HR~3222 {\it vs} HR~3222}

\figcaption{Radial distributions of stellar line-of-sight
velocities in $P.A.=153\arcdeg$ ({\it a}) and in $P.A.=145\arcdeg$
({\it b}); Gauss analysis has permitted us to extract three
kinematical stellar subsystems over the whole radius range,
the most intense contributor to the correlation peak at every
radius is presented by enhanced signs}

\figcaption{Isophotes of the surface brightness distributions
in the emission lines $H_\alpha$ ({\it a}) and [NII]$\lambda$6583
({\it b}) for the central part of NGC~2841. North is up, east is
to the left, the total size of the area shown is
$29\arcsec \times 29\arcsec$}

\figcaption{A map of [NII]$\lambda$6583-to-$H_\alpha$ flux ratio
(grey-scaled) with the superimposed $H_\alpha$ surface brightness
isophotes; the position angle of the top is 74\arcdeg\ (north is to
the right, east is up), the total sizes of the map are
$2\arcmin \times 2\arcmin$}

\figcaption{A piece of the Fig.1a in the radius range of
$-20\arcsec \div -80\arcsec$}

\figcaption{Results of the isophote analysis for the broad-band
$VRI$ images of the central part of NGC~2841: {\it a} ---
major axis position angle radial variations, {\it b} ---
ellipticity radial variations, {\it c} --- boxiness radial
variations; a fat straight line marks a proposed location
of the large-scale bar in NGC~2841}

\figcaption{A scheme of the observational phenomena in NGC~2841;
though the kinematical peculiarities (the gas on plane orbits
orthogonal to the bar and stellar LOSVD multi-component
structure) are observed mostly along the major axis, the
two-dimensional areas are shaded for a better impression}

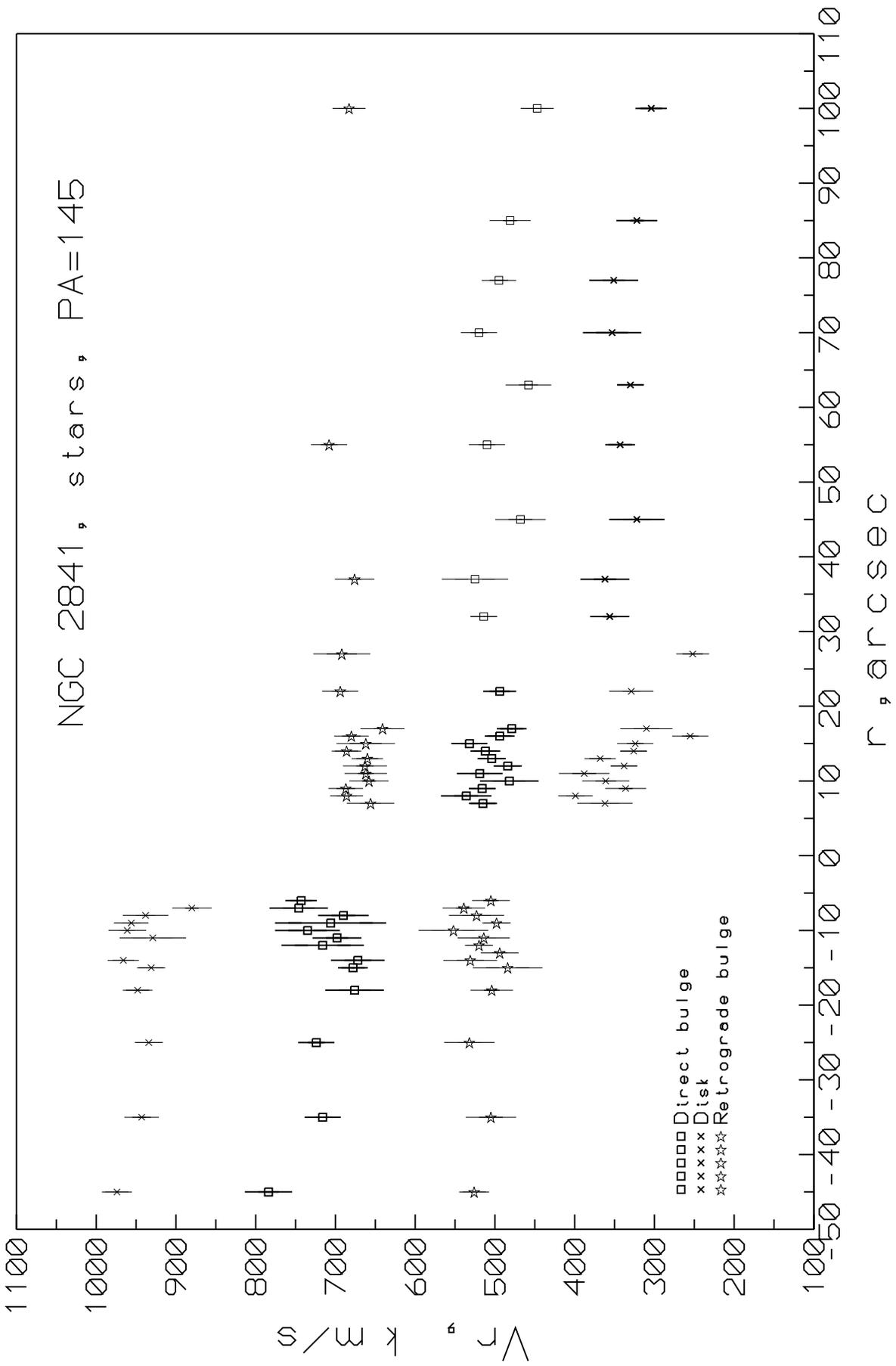
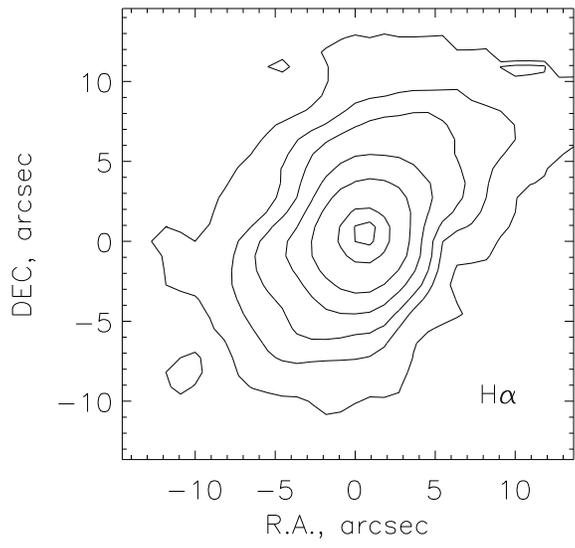
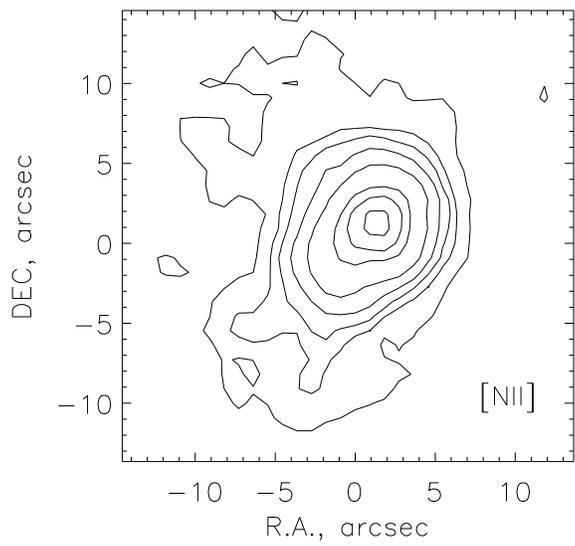
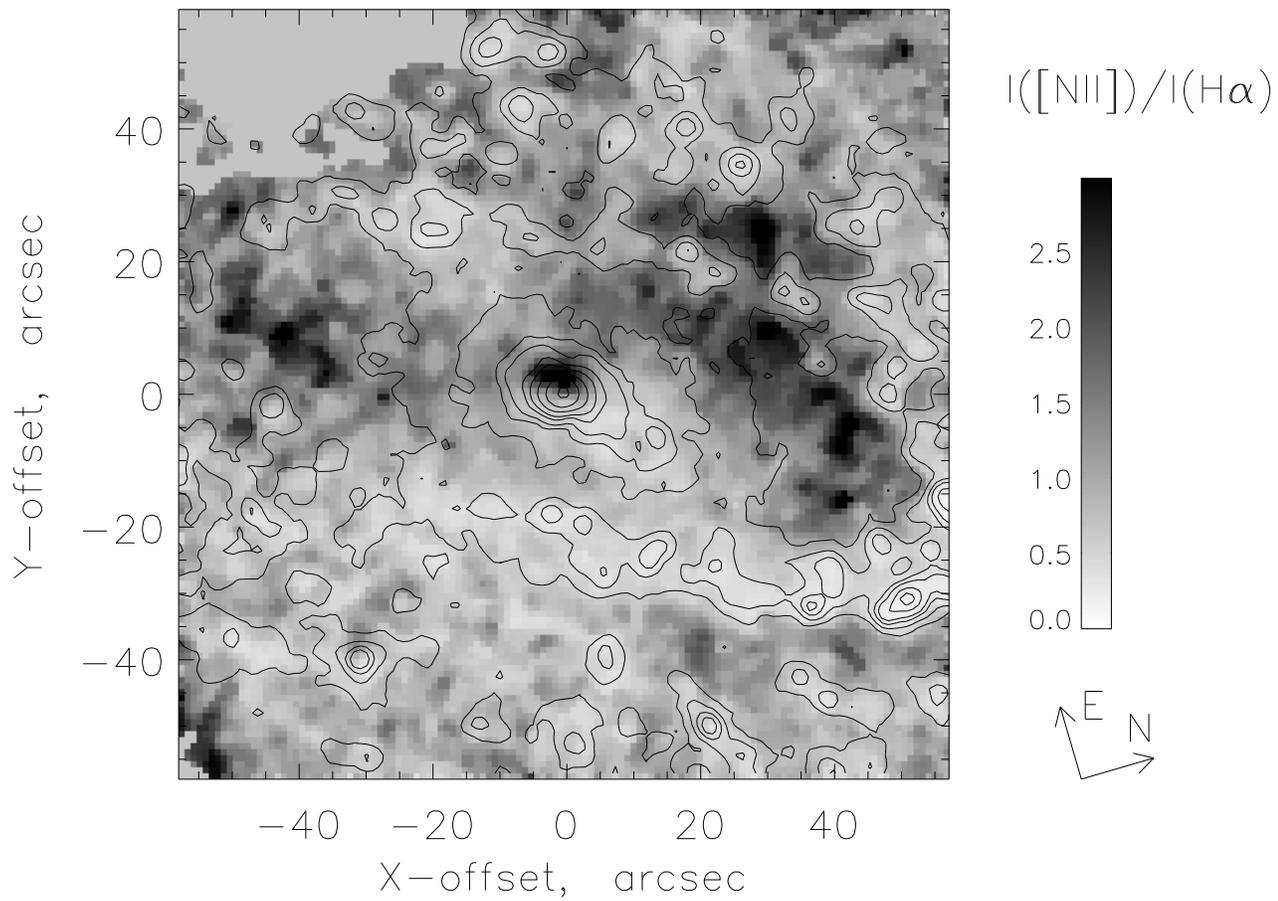
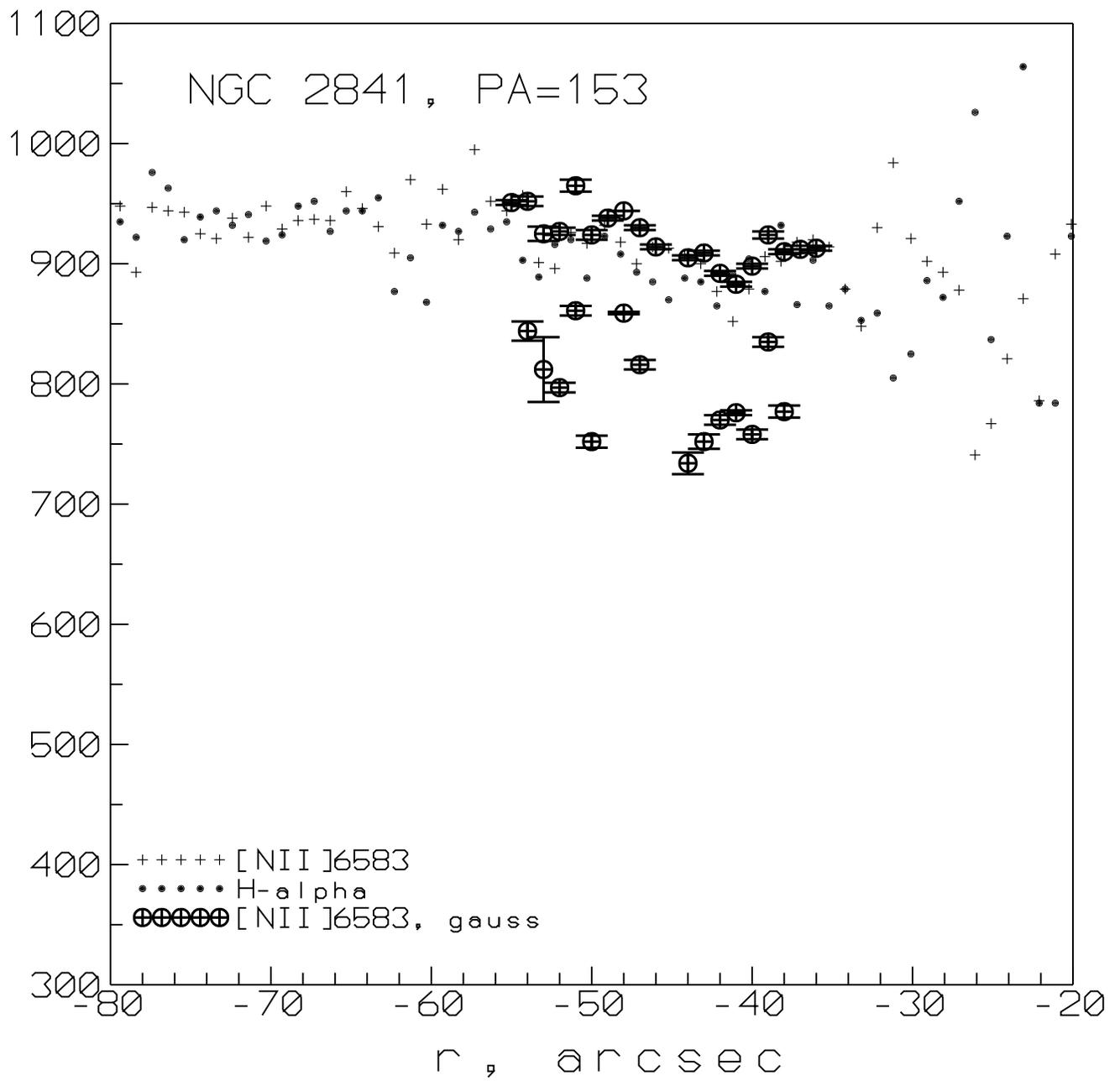
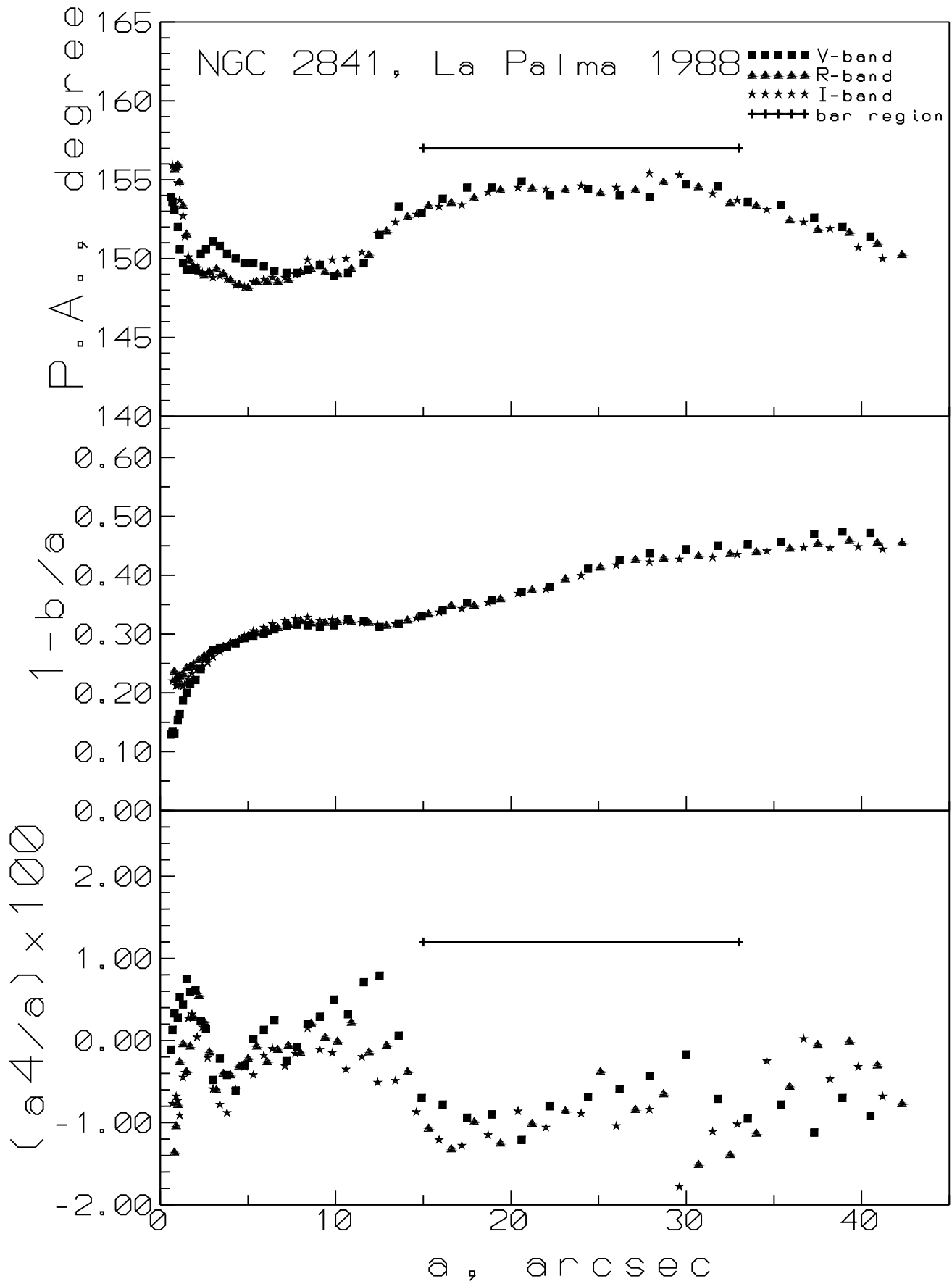
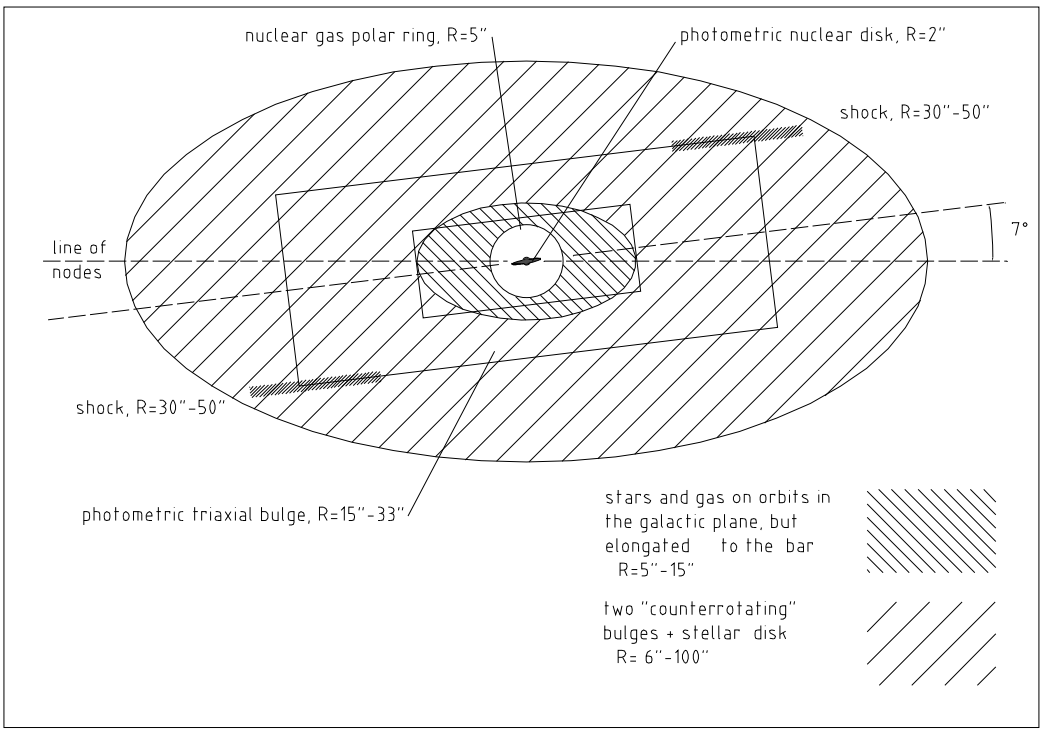
\end{document}